\documentclass[apj]{emulateapj}
\usepackage{epsfig}
\usepackage{amssymb}
\usepackage{amsmath}

\def\apj{{\it ApJ \,}}

\def\apjl{{\it ApJ} \,}

\def\mnras{{\it MNRAS} \,}

\def\aap{{\it A\&A} \,}

\def\pasj{{\it PASJ} \,}

\begin{document}

\title[DM and Galaxies Dynamics in A1689]{Quantifying the collisionless nature of dark matter and galaxies in A1689}

\author{
Doron Lemze\altaffilmark{1},
Yoel Rephaeli\altaffilmark{1,2},
Rennan Barkana\altaffilmark{1},
Tom Broadhurst\altaffilmark{1,3,4},
Rick Wagner\altaffilmark{2,5}
\& Mike L. Norman\altaffilmark{2,5}
}
\altaffiltext{1}
                {School of Physics and Astronomy, Tel Aviv University, Tel Aviv, 69978, Israel; doronl@wise.tau.ac.il}

\altaffiltext{2}
                {Center for Astrophysics and Space Sciences, University of California, San Diego, La Jolla, CA 92093, USA}

\altaffiltext{3}
                {Department of Theoretical Physics, University of Basque Country UPV/EHU,Leioa,Spain}

\altaffiltext{4}
                {IKERBASQUE, Basque Foundation for Science,48011, Bilbao,Spain}
\altaffiltext{5} {Physics Department, University of California, San Diego, La Jolla, CA 92093, USA}

\begin{abstract}

We use extensive measurements of the cluster A1689 to assess the
expected similarity in the dynamics of galaxies and dark matter
(DM) in their motion as collisionless `particles' in the cluster
gravitational potential. To do so we derive the radial profile of the 
specific kinetic energy of the cluster galaxies from the Jeans
equation and observational data. Assuming that the specific kinetic
energies of galaxies and DM are roughly equal, we obtain the
mean value of the DM velocity anisotropy parameter, and the DM
density profile. Since this deduced profile has a scale radius
that is higher than inferred from lensing observations, we tested 
the validity of the assumption by repeating the analysis using 
results of simulations for the profile of the DM velocity anisotropy. 
Results of both analyses indicate a significant difference between the 
kinematics of galaxies and DM within $r \lesssim 0.3r_{\rm vir}$. This 
finding is reflected also in the shape of the galaxy number density 
profile, which flattens markedly with respect to the steadily rising DM 
profile at small radii. Thus, $r \sim 0.3r_{\rm vir}$ seems to be a 
transition region interior to which collisional effects significantly 
modify the dynamical properties of the galaxy population with respect 
to those of DM in A1689

\end{abstract}


\section{Introduction}
\label{Introduction}

The properties of dark matter (DM), which is believed to be mostly 
cold and collisionless, have been extensively explored in dynamical 
simulations. A much touted result from these simulations is the 
``universal'' density profile (Navarro, Frenk, \& White 1997, 
hereafter NFW; Moore et al.\ 1998) of galaxies and galaxy clusters. 
As the main mass constituent of galaxy clusters, DM largely 
self-gravitates and dominates the hydrodynamics of intracluster (IC) 
gas and the dynamics of member galaxies. Cluster DM density profiles 
deduced from X-ray observations (Pointecouteau, Arnaud, \& Pratt 2005; 
Vikhlinin et al.\ 2006; Schmidt \& Allen 2007; Arnaud, Pointecouteau, 
\& Pratt 2008), galaxy velocity distribution (Diaferio, Geller, \& 
Rines 2005), SZ measurements (Atrio-Barandela et al.\ 2008), and
strong and weak lensing measurements (Broadhurst et al.\ 2005a,
hereafter B05a; Broadhurst et al.\ 2005b, hereafter B05b; Limousin et
al.\ 2007; Medezinski et al.\ 2007; Lemze et al.\ 2008, hereafter L08; 
Broadhurst et al.\ 2008; Zitrin et al.\ 2009, 2010; Umetsu et al.\ 2010)
are broadly claimed to be consistent with NFW profiles. However, the
shape of the profile in the inner cluster region where it is predicted
to have a characteristic radial slope of $-1$ 
has been deduced to be shallower in some studies (Kravtsov et al.\ 1998; 
Ettori et al.\ 2002; Sanderson et al.\ 2004) and steeper, around 
$-1.5$, in others (Fukushige \& Makino 1997, 2001, 2003; Moore et al.\ 
1999; Ghigna et al.\ 2000; Klypin et al.\ 2001; Navarro et al.\ 2004; 
Limousin et al.\ 2008).

A more complete description of the clustering properties of DM
necessitates characterization of its phase space distribution.
Whereas the DM density profile can be determined directly from
analysis of X-ray, lensing, and SZ measurements, deducing the DM
velocity distribution is considerably more challenging. On the other
hand, the dynamical properties of member galaxies can be studied
directly in terms of their density and velocity profiles, including
a study of the radial behavior of the velocity dispersion. Also, the
location of the velocity `caustics' 
can now be studied both in individual massive clusters where the data 
quality is high (Lemze et al.\ 2009, hereafter L09) and in composite 
surveys for which only lower quality measurements are available 
but statistical results can be derived (e.g., Biviano \& Girardi 2003).

The dynamical evolution of the galaxy population in a cluster is 
presumed to be largely collisionless following an initial phase 
of mean-field (`violent') relaxation of the main sub-cluster 
progenitors that merged to form the cluster. 
This collisionless behavior is expected particularly 
outside the central cluster region. As such, the basic dynamical 
characteristics of cluster galaxies are expected to resemble those 
of DM, which is strictly collisionless. For example, in the cluster 
1E0657-56 - the "bullet cluster" - despite a recent collision of 
two massive clusters, the spatial distributions of DM and galaxies
are quite similar. A conical shaped shock front is visible indicating 
two clusters have passed through each other with an obvious 
collisionally merged gas distribution, but the galaxies and the 
lensing mass are largely intact, implying straightforwardly that the 
DM and galaxies are collisionless (Markevitch et al.\ 2002; Clowe et 
al.\ 2004; Brada{\v c} et al.2006). In the colliding cluster A520, 
on the other hand, a massive dark core is claimed to coincide with 
the central X-ray emission peak, but the region is largely devoid of 
galaxies (Mahdavi et al.\ 2008), though this depends on the way the 
weak lensing analysis is formulated (Okabe \& Umetsu 2008).

With increasingly extensive and precise data, such as we have acquired
for A1689, it is now possible to assess the collisionless nature of
galaxies and DM by measuring the degree of consistency between the
measured galaxy and DM density profiles and the profile of the DM
velocity anisotropy. A1689 seems well relaxed (with possibly a small
deviation from a relaxed state; Andersson \& Madejski 2004). It has a
centrally located cD galaxy, and an X-ray emission region that is
spherically symmetric (Xu \& Wu 2002; L08; Riemer-Sorensen et al.\
2008). The cluster has well-defined galaxy velocity caustics with no
major infall of matter close to the virial radius (L09), and only a
low level of substructure (Broadhurst et al.\ 2005a,b; Umetsu \&
Broadhurst 2008).

We have previously determined the DM and gas density profiles in A1689 from 
a combined analysis of lensing and X-ray measurements (L08), using an
approach that we refer to as model-independent, since we did not
assume particular functional forms for the profiles. Additionally, we
extended the analysis by including photometric and spectroscopic
measurements of a very large number of galaxies in the A1689 field,
from which we deduced the positions and radial velocities of
$476^{+27}_{-43}$ cluster members. These results made it possible to
deduce the galaxy velocity anisotropy profile, which was found to exhibit 
the expected behavior, varying between predominantly radial orbits at 
large radii towards more tangential orbits near the center (L09).

Here we show that with the above information we can infer the DM
velocity anisotropy which we compare with the galaxy velocity anisotropy 
profile and with results from simulations. The paper is organized as follows; 
in \textsection~\ref{Methodology} we 
describe our method for determining the DM velocity anisotropy. In 
\textsection~\ref{Results} we compare the DM and galaxy density profiles 
(\textsection~\ref{Density profiles}), 
derive the DM velocity anisotropy and compare it with the galaxy 
velocity anisotropy profile and results from simulations 
(\textsection~\ref{Velocity anisotropy}), and estimate the 
collisionless profile of cluster galaxies (\textsection~\ref
{The collisionless profile of cluster galaxies}). We conclude with
a summary and discussion in \textsection~\ref{Discussion}.

\section{Methodology}
\label{Methodology}

In this section we present the procedure for deriving the DM velocity
anisotropy using results from our previous analyses 
of the galaxy dynamics (L09) and the total mass density profile (L08) 
of A1689. In L08 we combined lensing and X-ray measurements to 
determine model-independent profiles of the gas and total 
mass density profiles 
(i.e., without assuming particular functional forms). 
In the second stage of the work (L09) the galaxy surface number density 
and the projected velocity dispersion were included and analyzed using 
the Jeans equation, from which we obtained profiles of the 3D galaxy
number density and the galaxy velocity anisotropy.

The dynamics of a collisionless gas are governed by the Jeans equation
(Binney \& Tremaine 1987)
\begin{equation}
\frac{1}{\rho_{i}}\frac{d}{dr}\left(\rho_{i} \sigma_{i,r}^2  \right)
+\frac{2\beta_i \sigma_{i,r}^2}{r}= -\frac{GM}{r^2}\ ,
\label{Jeans eq}
\end{equation} 
where $i=$DM, gal, and $\rho_{i}(r)$ is the density of element $i$. The 
velocity anisotropy profile $\beta_i(r)$ is 
\begin{equation}
\beta(r)\equiv 1-\frac{\sigma_t^2(r)}{\sigma_r^2(r)}\ ,
\end{equation} 
where $\sigma_r(r)$ is the radial velocity dispersion, and 
$\sigma_t(r) = \sigma_{\theta}(r) = \sigma_{\phi}(r)$ is the (1D) 
transverse velocity dispersion. Using eq.~\ref{Jeans eq} for the 
galaxies, the degeneracy between $\sigma_{\rm gal,r}$ and 
$\beta_{\rm gal}$ can be removed with sufficient spectroscopic data 
(L09). This procedure is obviously irrelevant in the case of DM, for 
which we have to adopt an alternative approach.

The orbit of a test particle in a collisionless gravitational system
is independent of the particle mass. 
This would presumably imply that once hydrostatic equilibrium is 
attained, most likely as a result mixing and mean field relaxation, 
DM and galaxies should have the same mean specific kinetic energy, i.e.,
\begin{equation}
\sigma_{\rm DM,tot}^2(r)=\sigma_{\rm gal,tot}^2(r)\ , \label{sigma_total} 
\end{equation} 
where 
\begin{equation}
\sigma_{i,\rm tot}^2(r)= \sigma_{i,r}^2(r)+ \sigma_{i,\theta}^2(r)+ 
\sigma_{i,\phi}^2(r)=\sigma_{i,r}^2(r)\left(3-2\beta_i(r) \right)\ . \label{sigma_total derivation} 
\end{equation} 
Additionally, it is expected that the total specific kinetic energy of
DM particles is proportional to that of the gas (e.g., Mahdavi 2001;
Host et al.\ 2008), $T\propto \sigma_{\rm tot}^2$.  Observational
evidence for this scaling relation comes from combined X-ray and
optical observations of groups and clusters for which the mean
emission-weighted gas temperature scales roughly as the second power
of the total galaxy velocity dispersion (Mulchaey \& Zabludoff 1998;
Xue \& Wu 2000). 

The temperature and density profiles of IC gas can be deduced from
X-ray spectral and surface brightness measurements. These profiles can
then be used to determine the total mass distribution from a solution
of the hydrostatic equilibrium equation. We have recently developed a
model-independent joint lensing/X-ray analysis procedure 
(L08) to examine the consistency of X-ray temperature and emission 
profiles with the lensing based mass profile, finding that the cluster 
temperature profile is systematically $\sim$ 30-40\% lower than expected 
when solving the equation of hydrostatic equilibrium using the
lensing-deduced mass profile. This discrepancy may reflect in part the
ambiguity in deriving 3D temperatures from projected spectral 
X-ray data, stemming from the sampling of a range of gas temperatures 
along any given line of sight (Mazzotta et al.\ 2004; Vikhlinin 2006). 
This could also be partly 
related to the small-scale structure of the gas [possibly including 
relatively dense cooler clouds found in simulations (Kawahara et al.\ 
2007)] which may result in a significant downward bias in 
temperature estimates from spectral X-ray observations. 
The inferred temperature is also sensitive to instrumental effects, as 
has recently been deduced in the analysis of Chandra observations of 
A1689 (Peng et al.\ 2009). Other reasons for temperature biases can be 
deviations from equilibrium and non-thermal pressure (Molnar et al.\ 
2010). Even in the best-case scenario the temperature can be used as a 
reliable tracer for $\sigma_{\rm tot}$ only in the inner part of the 
cluster, where hydrostatic equilibrium applies, and also only at radii 
larger than about $0.1r_{\rm vir}$, since at smaller radii the specific 
energy of the gas and DM may be different (Rasia, Tormen, \& Moscardini 
2004).

For our analysis we assume an NFW-like profile for the DM density 
\begin{equation}
\rho_{\rm DM}(r)\propto [(r/r_s)(1+r/r_s)^{\alpha}]^{-1}\ ,
\end{equation} 
and take the DM velocity anisotropy to be constant, i.e., $\beta_{\rm
DM} ={\rm const}$, since the data is not sufficient to meaningfully
constrain more than one free parameter in this quantity. We then 
relate our model parameters to the DM radial velocity dispersion,
$\sigma _{\rm DM,r}$, using the Jeans equation, where the (total) mass
profile is taken from L08. Here we allow for the possibility of a
difference between the total density profile (directly measured by
lensing) and the profile of just the DM. From the DM velocity
anisotropy and radial velocity dispersion we can then deduce the DM
total specific kinetic energy. Best-fit values of $\rho _{\rm DM}(r)$
and $\beta_{\rm DM}$ are determined by fitting the DM total specific
kinetic energy to the galaxy total specific kinetic energy. More
specifically, $\beta_{\rm DM}$ and $\rho_{\rm DM}(r)$ are determined
by minimizing $\chi^2=
\sum_i V^{T}(r_i)\cdot C^{-1}
\cdot V(r_i)$, where $V(r_i) = \sigma_{\rm DM,tot}^2(r_i)-\sigma_
{\rm gal,tot}^2(r_i)$ and C is the covariance matrix of the measured
$\sigma_{\rm gal,tot}^2(r_i)$. The galaxy total specific kinetic
energy itself is not a direct measurement. It was derived as shown in
eq.~\ref{sigma_total derivation} from $\sigma_{\rm gal,r}$ and
$\beta_{\rm gal} (r)$. Since the values of $\sigma_{\rm gal,tot}$ at
the various radial positions $r_i$ are derived from underlying
parameterized expressions (as given below for $\beta_{\rm gal} (r)$),
their uncertainties are correlated. 

The derivation of $\beta_{\rm gal}(r)$ and $\sigma_ {\rm gal,tot}(r_i)$ 
is carried out by following the same procedure as 
in L09, except for allowing a higher freedom of $\beta_{\rm gal}(r)$ at 
large radii. N-body simulations for a variety of cosmologies show that 
the velocity anisotropy has a nearly universal radial profile (Cole \& 
Lacey 1996; Carlberg et al.\ 1997). In accord with the work of L09, 
$\beta_{\rm gal}(r)$ is taken to have the following form:
\begin{equation}
\beta_{\rm gal}(r) = (\beta_0+\beta_{\infty})\frac{(r/r_c)^2}{(r/r_c)^2+1}-\beta_0\ , 
\label{beta analytic expression}
\end{equation} 
where we note that $\beta_{\infty}=1$ was adopted by L09. 
Now, the total number of $\sigma_{\rm gal,tot}(r_i)$ bins cannot exceed 
the total number of free parameters in the expressions for $\sigma_{\rm 
gal,r}$ and $\beta_{\rm gal}(r)$, which is $6$, since a larger number 
would cause a complete degeneracy among the various values of 
$\sigma_{\rm gal,tot}(r_i)$. Even taking $6$ bins resulted in 
an unphysical correlation matrix (i.e., one having a negative 
eigenvalue), which still indicates 
a near-degeneracy. Therefore we adopted 5 radial bins of 
$\sigma_{\rm gal,tot}(r_i)$, which was the maximum number 
for which degeneracy is not significant and error estimates are 
reasonable.

\section{Results}
\label{Results}

\subsection{Velocity anisotropy profiles}
\label{Velocity anisotropy}

\begin{figure}
\centering
\epsfig{file=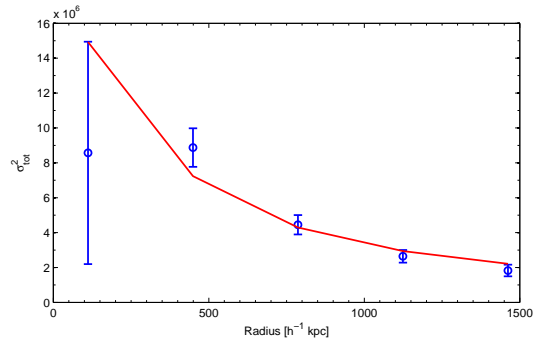, width=8cm, clip=}
\caption{Profile of the total specific kinetic energy 
of the galaxies (blue circles and 1--$\sigma$ errorbars) and the 
fitted total specific kinetic energy of the DM (red line).
\label{sigma_tot_fit}
}
\end{figure}

The DM velocity anisotropy, $\beta_{\rm DM}$, was determined 
as described in \textsection~\ref{Methodology}, with 
the constraint that the DM total specific kinetic energy must satisfy 
$\sigma_{\rm DM,tot}^2\geq 0$. The resulting acceptable fit, 
with $\chi^2/{\rm dof}=3.5/(5-3)$, is shown in figure~\ref{sigma_tot_fit}.
The best-fit parameters of the analytical expressions for the DM
density and velocity anisotropy profiles are given in
table~\ref{best-fit parameters table}. It is important to note that
while the errors on the parameters are rather large, in this fit we
have constrained the DM parameters based only on the fit to
equation~\ref{sigma_total}, without assuming the DM density profile to
be similar to the total mass density profile.  Thus, the results allow
a largely independent comparison between the consequences of assuming
equation~\ref{sigma_total} and the results of other observational
probes of the cluster.
\begin{table}
\caption{The values of the parameters of DM density and velocity anisotropy. 
The errors are 1-$\sigma$ confidence. \label{best-fit parameters table}}
\begin{center}
\begin{tabular}{|l|c|}

\hline
Parameter & Value \\
\hline                    
 $r_s$ \;\;\; [h$^{-1}$ kpc] & $1330_{-605}^{+1210}$            \\
 $\alpha$          & $2.79_{-0.76}^{+1.27}$           \\
 $\beta_{\rm DM}$           & $0.49_{-0.27}^{+0.13}$           \\ 
\hline
\end{tabular}
\end{center}
\end{table}

\begin{figure}
\centering
\epsfig{file=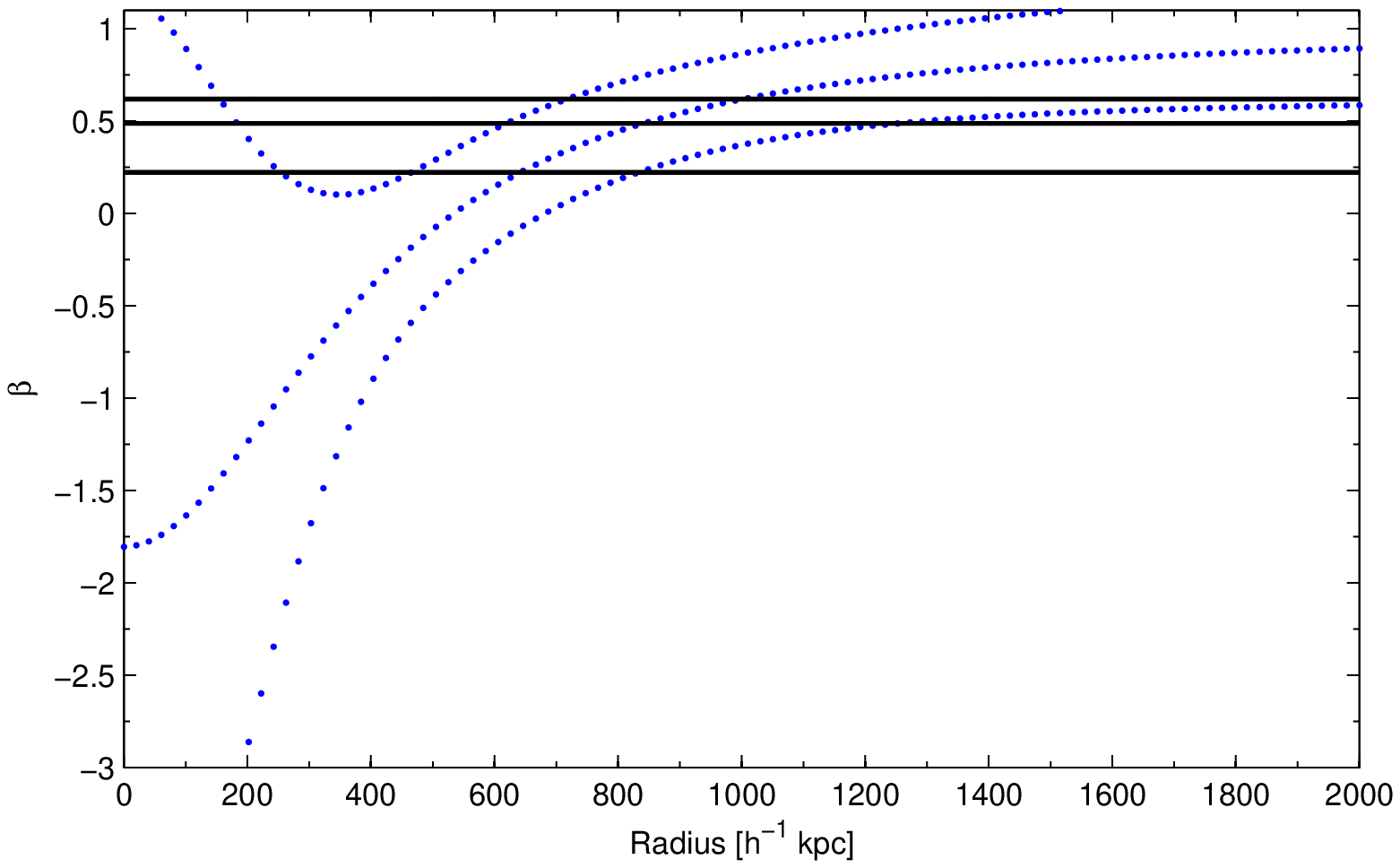, width=8cm, clip=}
\epsfig{file=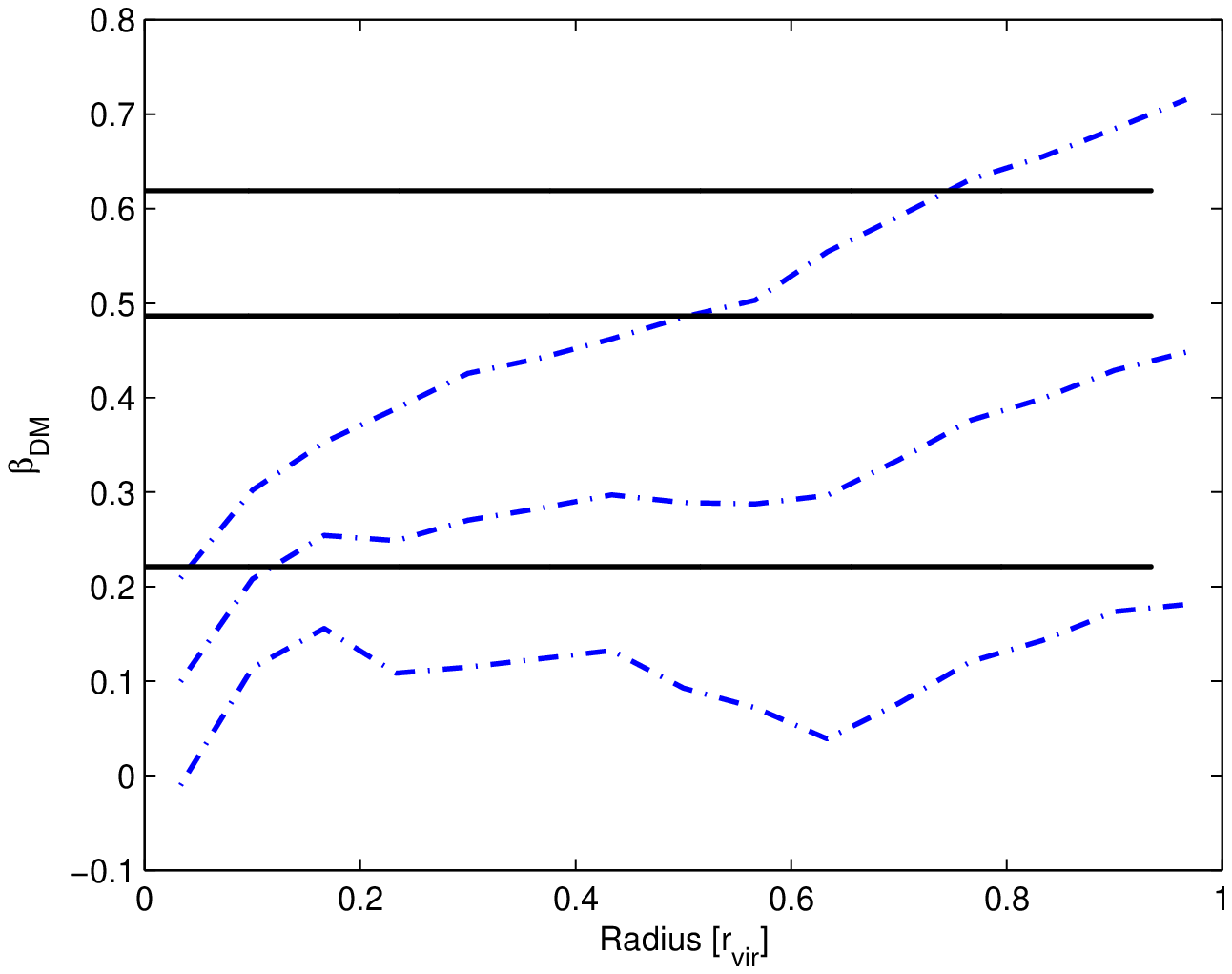, width=8cm, clip=}
\caption{Top panel: comparison between $\beta_{\rm DM}$ and $\beta_{\rm 
gal}$. Shown are the velocity anisotropy profiles of galaxies (blue dotted 
lines) and DM 
(black solid lines). Bottom panel: comparison between the DM velocity 
anisotropy inferred from data (black solid lines) and that derived 
from simulations (blue dot-dashed lines). For each set 
the central line shows the best fit and the other two 
lines show the $\pm$ 1--$\sigma$ uncertainty range.
\label{beta_profile_DM_Vs_gal}
}
\end{figure}

The corresponding galaxy and DM velocity anisotropy profiles are
plotted in figure~\ref{beta_profile_DM_Vs_gal} (top panel) together
with their respective 1--$\sigma$ uncertainties.  In
figure~\ref{beta_profile_DM_Vs_gal} (lower panel) we compare the
derived DM velocity anisotropy value to the profile derived from
simulations. The current data only allow us to determine an overall,
typical value of $\beta_{\rm DM}$ in the cluster, and given the rather
large uncertainties, there is fair agreement between this value and
the typical value of $\beta_{\rm gal}$ in this cluster, and also with
the typical values of $\beta_{\rm DM}$ seen in simulated clusters.
Note from the figure that while we allowed $\beta_{\infty}$ to be a
free parameter for the galaxies, the best-fit value came out quite
close to unity.

We briefly describe the derivation of the DM velocity anisotropy profile 
from results of an ENZO simulation (for a more complete description of this 
particular simulation, please refer to Hallman et al.\ 2007). 
Since A1689 is a moderately-distant (z=0.183), high mass cluster, 
$M_{\rm vir} \sim 1.5\cdot10^{15}$ h$^{-1}$ M$_{\odot}$ (Broadhurst 
et al.\ 2005a; Oguri et al.\ 2005; Limousin et al.\ 2007; L08; Umetsu 
\& Broadhurst 2008; L09; Umetsu et al.\ 2009; Corless et al.\ 2009; 
Coe et al.\ 2010), we derive the DM velocity anisotropy profiles 
for a sample of high-mass clusters with $M_{\rm vir} 
> 10^{15}$h$_{0.7}^{-1}$ M$_{\odot}$ at $z=0.2$. The clusters were drawn 
from a cosmological adaptive mesh refinement (AMR) simulation performed 
with the ENZO code developed by Bryan \& Norman (1997) and Norman \& 
Bryan (1999) assuming a spatially flat CDM model (very similar to the 
concordance model). The AMR simulation assumed adiabatic gas dynamics 
(i.e., neither radiative heating, cooling, star formation or feedback 
were included). The box size was 512 h$^{-1}$ Mpc comoving on a side 
with $512^3$ DM particles, and DM mass resolution of about $10^{11}$ 
h$_{0.7}^{-1}$ M$_{\odot}$. The root grid contained $512^3$ grid cells, 
and the grid was refined by a factor of two, up to seven levels, 
providing a spatial resolution of $6.5$ kpc h$^{-1}$ at $z=0.2$. For 
each halo out of the total number of 51, we derive the DM velocity 
anisotropy profile, which is averaged over all halos, with an uncertainty that 
is taken to be the standard deviation as determined by Lemze et al.\ 
(2010, in preparation). The derived profile is in agreement with those 
found in other studies (e.g., Colin, Klypin, \& Kravtsov 2000; DM04; 
Mamon \& Lokas 2005; Valdarnini 2006).

\subsection{Density profiles}
\label{Density profiles}

\begin{figure}
\centering
\epsfig{file=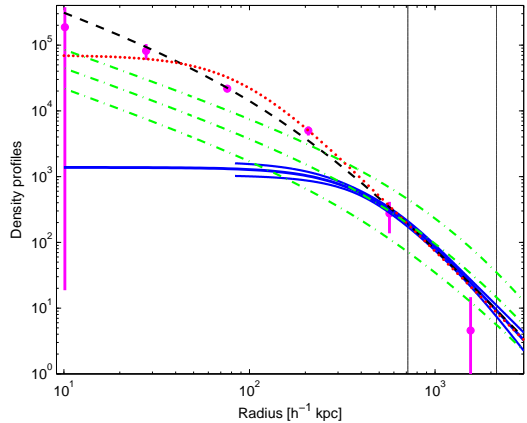, width=8cm, clip=}
\caption{The galaxy density (blue solid curve with upper and lower
solid curves marking the 1--$\sigma$ uncertainty) is compared to the
deduced DM density (green dot-dashed curves showing the best-fit and
and 1--$\sigma$ uncertainty) and to the total matter density for the 
model-independent fit (points with error bars), the NFW fit (black 
dashed curve), or the core fit (red dotted curve). We show 
{\em relative}\/ density profiles, all scaled to match at 700 h$^{-1}$ 
kpc (except for the galaxy 1--$\sigma$ lines). The left and
right black vertical lines indicate $\frac{1}{3} r_{\rm vir}$ and
$r_{\rm vir}$, respectively. Note that the 
galaxy density line at low radii is an extrapolation due to lack of 
data in this region (and therefore no error bars are shown). 
\label{density_profiles_comparison}
}
\end{figure}

To further explore phase space occupation of DM and galaxies, 
we compare the deduced DM density profile to the galaxy and 
to the total density profile. The shape of the galaxy density profile is 
represented with a core, and that of the total density with either a
model-independent profile, a `universal' NFW form, or a cored form. In
figure~\ref{density_profiles_comparison} we compare our results for
these profiles, i.e., we compare the DM density from our current fitting to 
equation~\ref{sigma_total} (green dot-dashed curves showing the 
best-fit result as well as the 1--$\sigma$ uncertainty
region) to our previous results for the galaxy density (blue solid
curves) and the total mass profile, whose various versions are shown
by the points with error bars (model-independent fit), dashed black
curve (NFW), and dotted red curve (core). In order to include all these 
in the same figure, and allow a comparison of the relative shapes,
we arbitrarily scaled the profiles so that they match at 700 h$^{-1}$
kpc $\sim\frac{1}{3}r_{\rm vir}$ (with arbitrary units in the $y$-axis).

While the DM scale radius derived here from the velocity dispersion
fit has a large error, its best-fit value is significantly higher than
that derived previously (L08, table 4) for the total (mostly DM) density. 
This is reflected by the slower rise of the DM density at small radii 
compared to the total density 
(figure~\ref{density_profiles_comparison}). This may indicate that our
assumption of eq.~\ref{sigma_total} invalid at small radii.

\begin{figure}
\centering
\epsfig{file=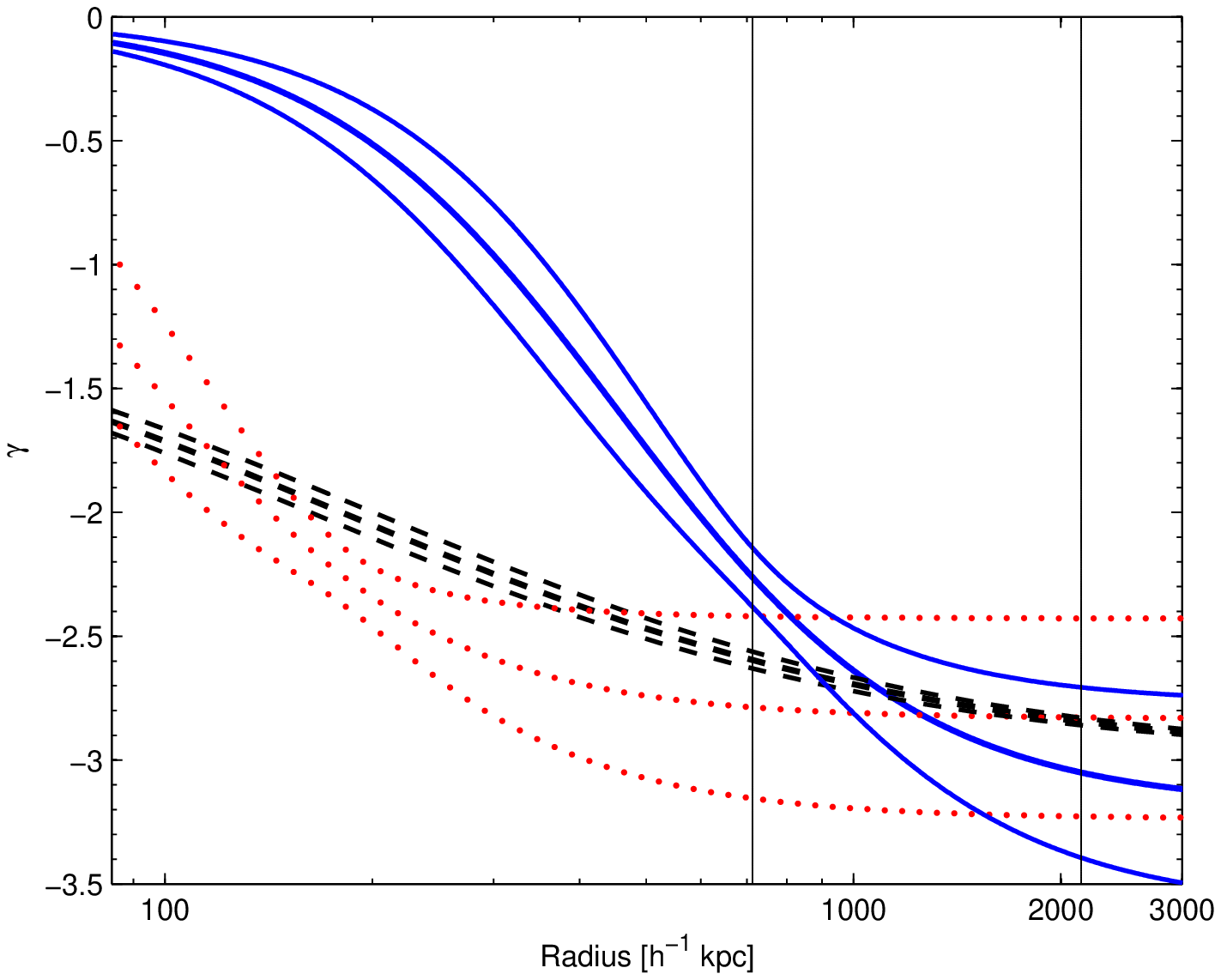, width=8cm, clip=}
\epsfig{file=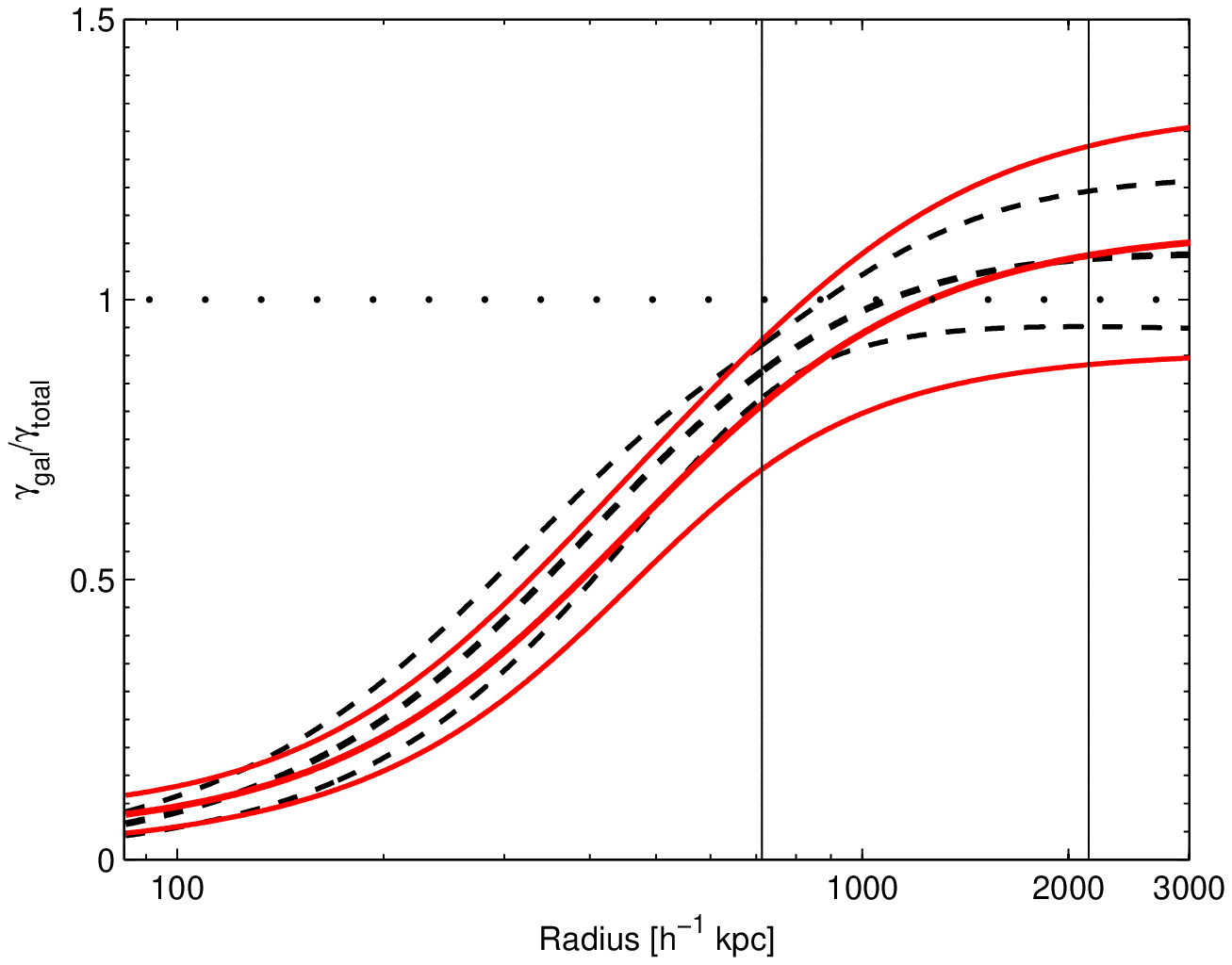, width=8cm, clip=}
\caption{Top panel: power-law indices of the galaxy and total mass 
profiles. The index of the galaxy profile is shown by the blue solid 
line, with upper and lower solid lines indicating 1--$\sigma$ 
uncertainty, and that of the total mass is shown by the black dashed 
(NFW) and dotted red (core) lines, with 
1--$\sigma$ uncertainties. Bottom panel: ratio between the power-law
index of the galaxy density to that of the total matter (NFW - black
dashed with upper and 1--$\sigma$ uncertainty; core - red solid with 
1--$\sigma$ uncertainty). The left and right black vertical lines 
mark $\frac{1}{3} r_{\rm vir}$ and $r_{\rm vir}$, respectively.
\label{gamma_comparison}
}
\end{figure}

Another perspective on the assumption that 
DM follows the same dynamics as the galaxies can be  
seen from the steepness of the directly measured density profiles. To assess the 
steepness of the fitted profiles we plot in 
figure~\ref{gamma_comparison} the distribution of their power-law 
indices,
\begin{equation}
\gamma(r)=\frac{d\log[\rho(r)]}{d\log[r]}\ .
\end{equation}
The power-law index of the galaxy profile is shown by the blue lines,
and that of the total mass profile by the black (NFW) and red (core)
lines. We did not plot also 
the power-law slope of the model-independent fit, since it is clear 
from figure~\ref{density_profiles_comparison} that it would look very
similar to the results for the NFW and core profiles. Since the DM 
profile should be rather similar to the total mass profile, 
the conclusion from these figures here again is 
that the galaxies and the DM have consistent density profiles for $r \ga
\frac{1}{3} r_{\rm vir}$, but the profiles are significantly different
at smaller radii.

\subsection{The collisionless profile of cluster galaxies}
\label{The collisionless profile of cluster galaxies}

In \textsection~\ref{Velocity anisotropy} we derived the DM velocity 
anisotropy profile assuming that both galaxies and DM are collisionless, 
finding that the best-fit value of DM density scale radius is higher 
than inferred from direct lensing observations for the total mass 
(L08). As was mentioned above, this may indicate that 
eq.~\ref{sigma_total} is not valid at all radii. To quantify differences 
between $\sigma_{\rm DM,tot}^2$ and $\sigma_{\rm gal,tot}^2$ we use the 
ratio 
\begin{equation}
 f_{\rm coll}(r)\equiv\frac{\sigma_{\rm DM,tot}^2(r)}{\sigma_{\rm gal,tot}^2(r)}\ ,
\end{equation}
which we expect to be very close to unity 
if both components are fully 
collisionless (or if both deviate comparably from the purely 
collisionless limit).
 
\begin{figure}
\centering
\epsfig{file=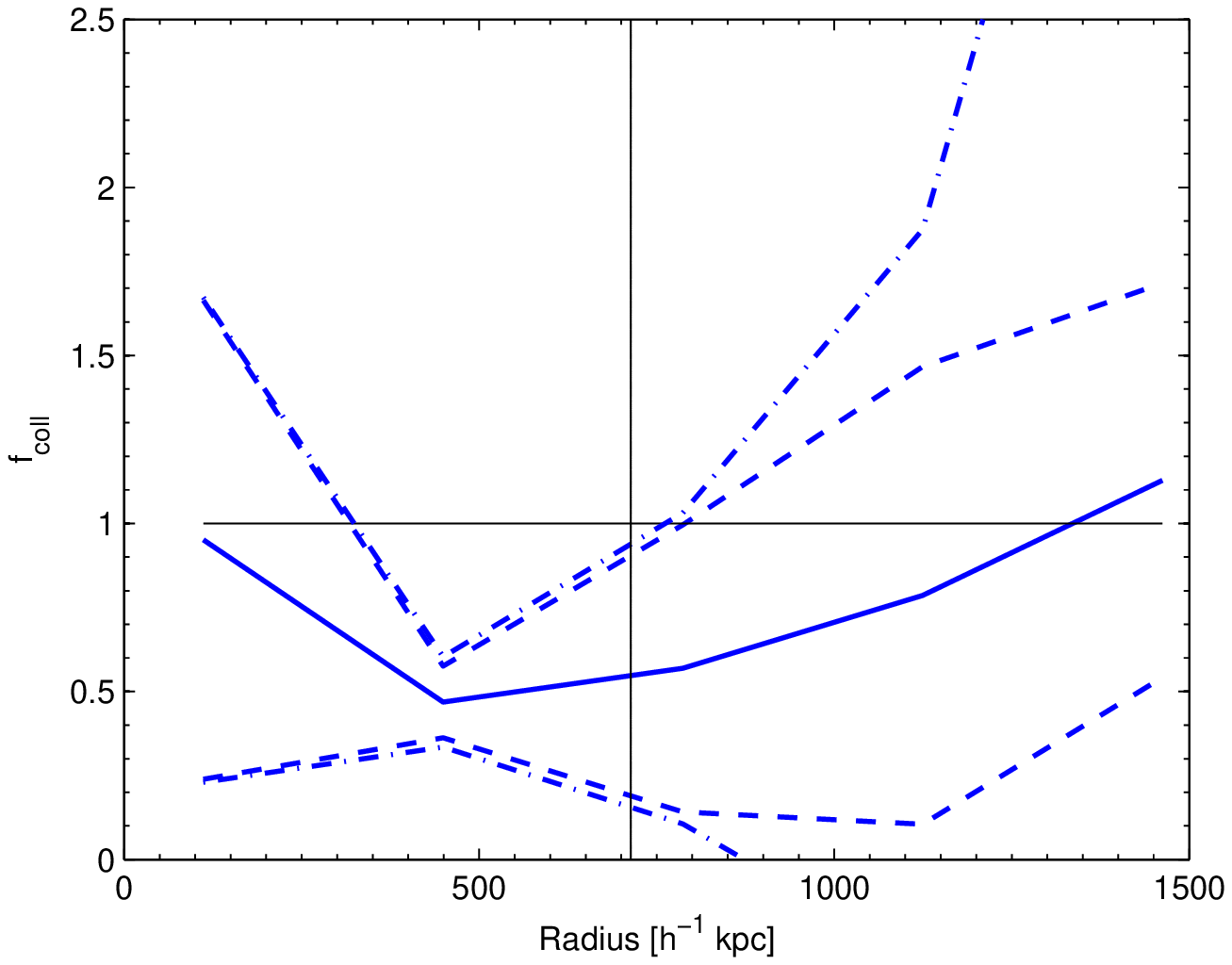, width=8cm, clip=}
\epsfig{file=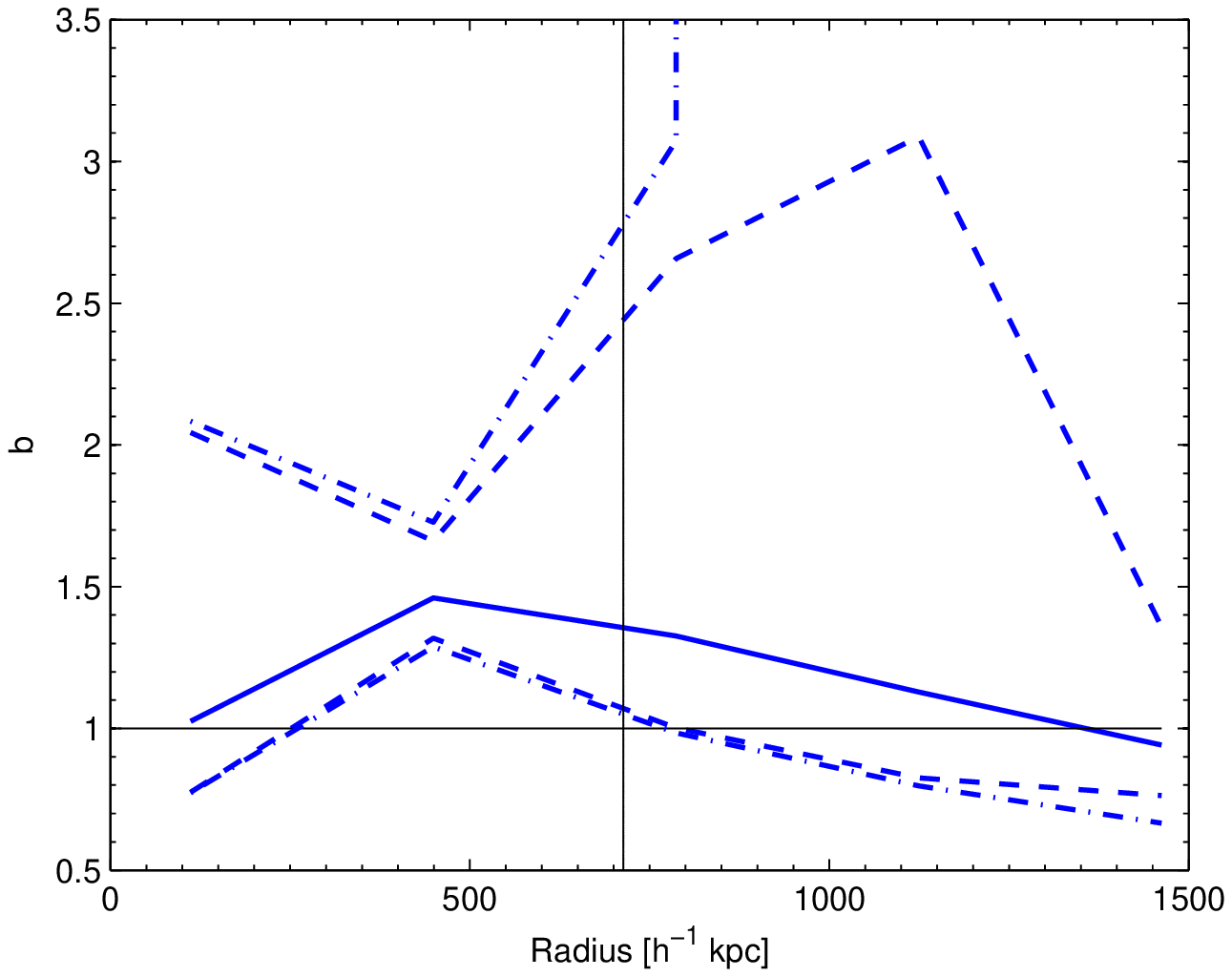, width=8cm, clip=}
\caption{Profiles of $f_{\rm coll}$ (top panel, blue solid curve) 
and the velocity bias $b$ (lower panel, blue solid curve) taking the
total and gas density model independent profiles from L08 along with
their 1--$\sigma$ uncertainty regions (marked by 
blue dot-dashed and dashed lines, for the case when the uncertainty of 
DM velocity anisotropy from simulations is included, and the case when 
this uncertainty is not included, 
respectively). The vertical black line marks 
$\frac{1}{3} r_{\rm vir}$, and the horizontal dotted line indicates 
the expected value if both DM and galaxies are purely collisionless.
\label{f_coll_model_independent}
}
\end{figure}

\begin{figure}
\centering
\epsfig{file=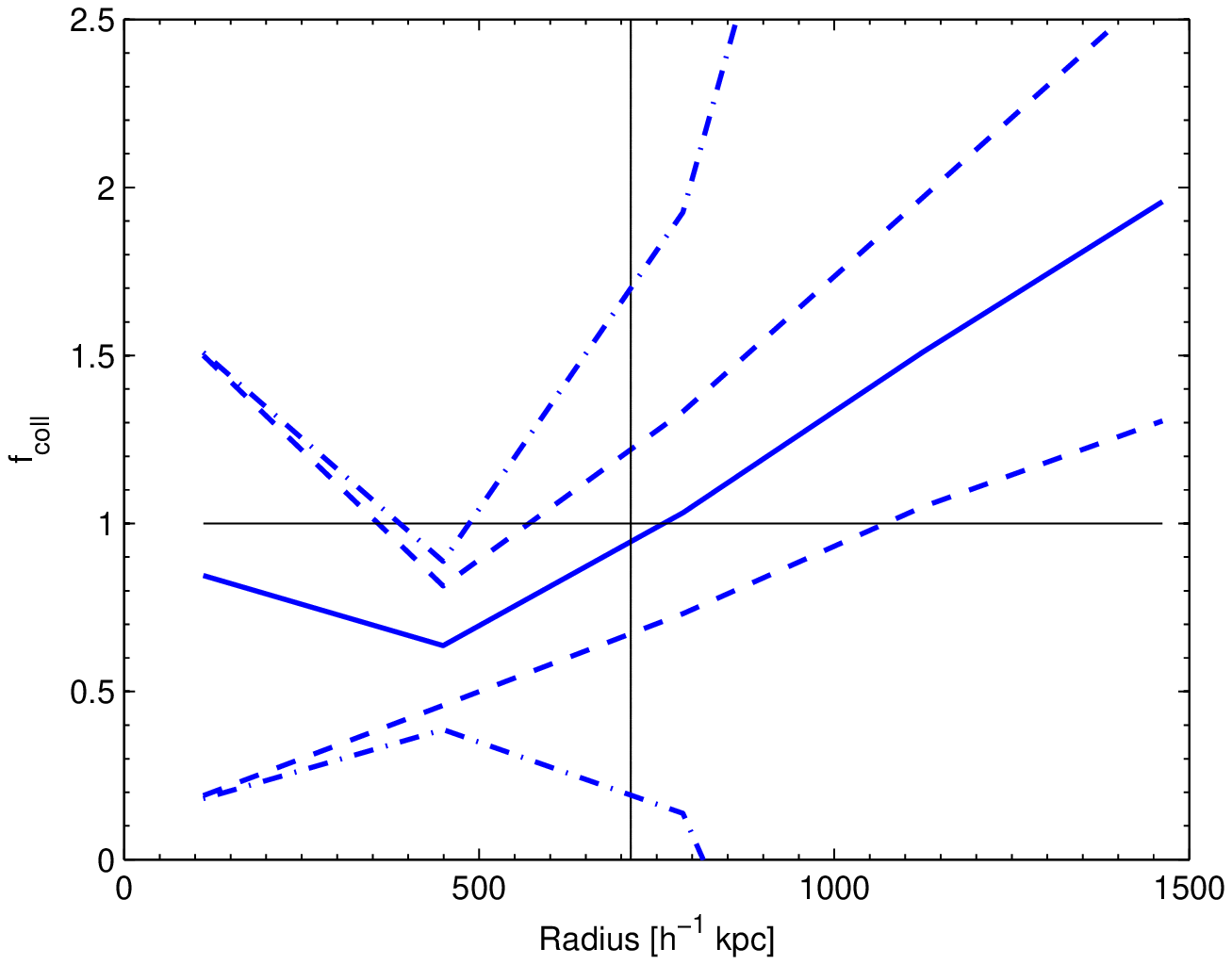, width=8cm, clip=}
\epsfig{file=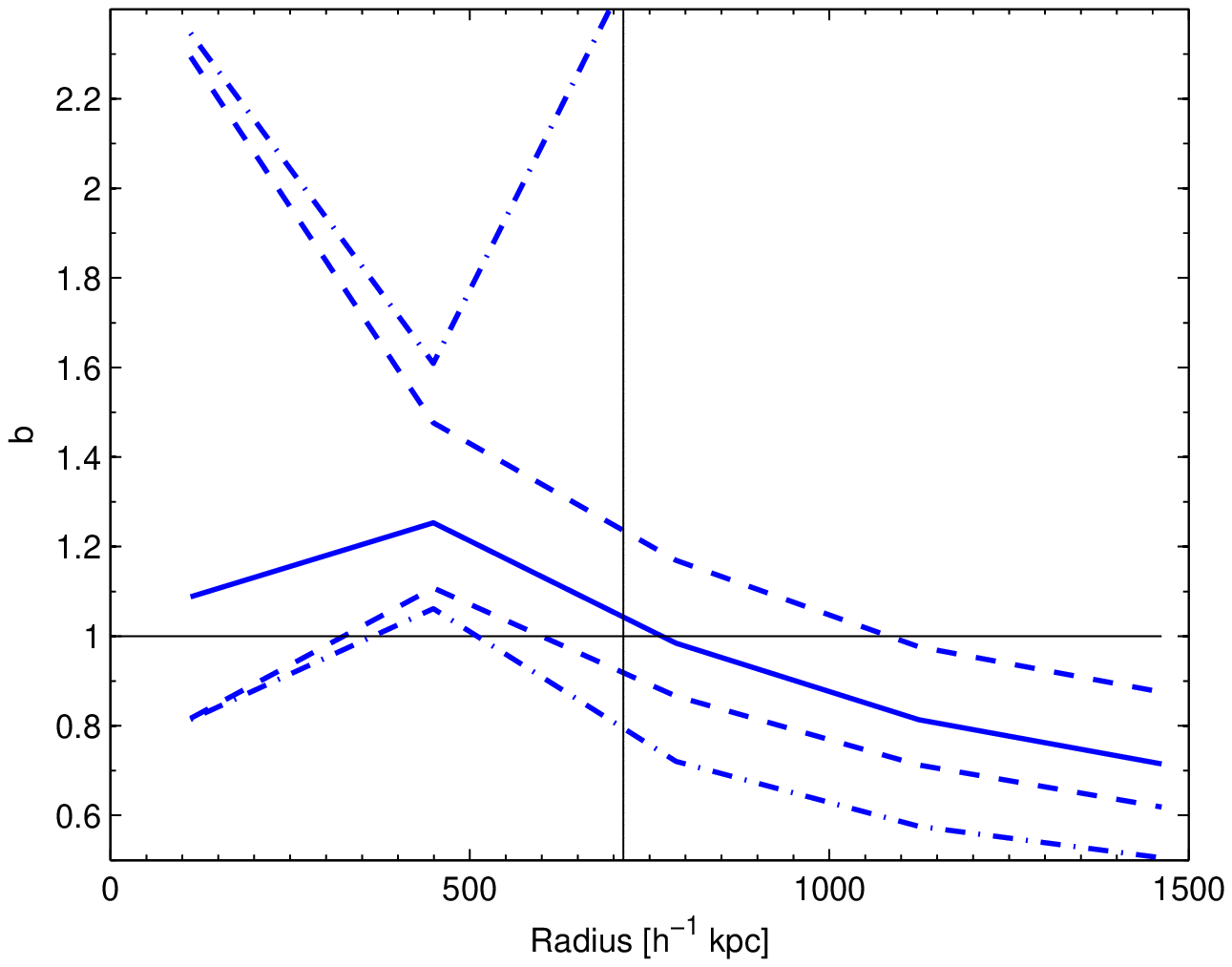, width=8cm, clip=}
\caption{The same as figure~\ref{f_coll_model_independent} 
except that here 
we take the total and gas density profiles to be NFW and double beta 
model (as described in L08), 
respectively. 
\label{f_coll_model_dependent}
}
\end{figure}

We have thus far assumed that $f_{\rm coll}(r)$ must equal unity and tried
to obtain acceptable fits under this assumption, 
deriving other results in the process. Here we try a different approach, 
where we start by assuming that the DM velocity anisotropy profile in 
A1689 matches the profile derived from simulations (see 
\textsection~\ref{Velocity anisotropy}). We used our 
previously-determined DM mass density, $\rho_{\rm DM} \simeq
\rho_{\rm tot}-\rho_{\rm gas}$, derived by assuming profiles for
the total and gas density profiles. 
Two kinds of profiles were assumed, a model-independent and a 
model-dependent one, 
where in the model-dependent case we assumed an NFW and double beta 
model for the total and gas density profiles, respectively.  After 
assuming a profile for the total and gas density profile, we 
can then solve the Jeans equation to determine $\sigma_{\rm DM,tot}^2$.  In
figures~\ref{f_coll_model_independent} and
\ref{f_coll_model_dependent} we show the resulting ratio $f_{\rm
coll}(r)$, or equivalently the corresponding velocity bias
of the galaxies relative to the DM
\begin{equation} 
b(r)=\sqrt{1/f_{\rm coll}(r)}\  , 
\label{f_coll b connection} 
\end{equation}
for the model-independent and model dependent profiles, respectively.
Our results are consistent with both the galaxies and the DM being
purely collisionless at $r\gtrsim \frac{1}{3} r_{\rm vir}$, but
there is some evidence for a significant deviation from this limit at
smaller radii (except for the very smallest radii, where the 
observational constraints are weak).


\section{Discussion}
\label{Discussion}

The work reported here is a continuation of our comprehensive study 
of the dynamical properties of DM and galaxies, and the hydrodynamical 
properties of IC gas in the well-observed cluster A1689 (L08, L09). 
In L09 we derived the galaxy density and velocity distributions, from 
which we deduced the specific kinetic energy of the galaxies. In the work reported here we 
assumed that if DM and galaxies are fully collisionless they should have 
the same average specific kinetic energy (as manifested in eq.~\ref{sigma_total}). 
Using the mass profile which was previously derived in L08, together with 
the Jeans equation, we fitted the DM to the galaxy specific kinetic 
energy and determined the DM density and the DM velocity anisotropy. 

For the DM density we obtained 
a best-fit value for the power law index at large radii, $\alpha+1$, 
which is higher (i.e., steeper) than the NFW value of 3 by about 
$1\sigma$.  
The best-fit value of the scale radius is $\simeq 1.9\sigma$ 
higher than the scale radius of the total density 
This was the first of several indications that the dynamical state of 
the galaxies and the DM may differ at small radii, below $\sim 
\frac{1}{3} r_{\rm vir}$. In particular, our assumption of 
eq.~\ref{sigma_total} resulted in a derived DM density profile at small 
radii that was inconsistent with (specifically, shallower than) that we 
have previously measured for the total matter density.  

In order to quantify the possible difference between $\sigma_{\rm
DM,tot}^2$ and $\sigma_{\rm gal,tot}^2$, we used another approach to
determine their ratio $f_{\rm coll}$. Specifically, we adopted the
profile of $\beta_{\rm DM}$ deduced from simulations and our derived
DM mass density (from L08), together with 
the Jeans equation. The deduced profile of $f_{\rm coll}$ can be 
interpreted as a measure of how collisionless are the galaxies with 
respect to the DM (which is equivalently expressed also by $b$, the 
velocity bias of eq.~\ref{f_coll b connection}). Note that in some 
simulation work 
the velocity bias is defined in terms of the velocity dispersions of
DM subhalos (rather than galaxies) compared to the DM background. We
find that $f_{\rm coll}<1$ (at $> 1\sigma$) at some radii below
$\frac{1}{3} r_{\rm vir}$, but that it is consistent with unity at
larger radii. This implies 
that at $r\gtrsim \frac{1}{3} r_{\rm vir}$ we can indeed assume 
the validity of eq.~\ref{sigma_total}, 
but at $r\approx 450$ h$^{-1}$ kpc we deduce $b \approx 
1.5^{+0.3}_{-0.2}$ and $b \approx 1.3^{+0.4}_{-0.2}$, for the
model-independent and model dependent profiles, respectively, an
indication that $b$ increases from large radii towards the center. This 
is in accord with results from simulations which found 
$b$ to vary from slightly higher than one (by $\sim10\%$) at large radii
(Ghigna et al.\ 2000; Colin, Klypin, \& Kravtsov 2000; 
Diemand, Moore, \& Stadel, \ 2004, hereafter DMS04; 
Gill et al.\ 2004) to $b \simeq 1.3$ in the central region ($r<0.3r_{\rm 
vir}$) (Ghigna et al.\ 2000; DMS04).

We note that there is some uncertainty regarding the dependence of
this bias on the features of the subhalos or galaxies considered; e.g.,
whereas DMS04 found the bias to be independent of the subhalo mass,
Ghigna et al.\ (2000) claim that there is no significant bias when
only high-mass subhalos are considered.  In an earlier work Diaferio
et al.\ (1999) found a bias for blue galaxies of about $1.5-2$, but no
bias for red galaxies. Their interpretation was that the bias is due
to the fact that blue galaxies are not in equilibrium. The velocity
bias can be explained by the fact that slow subhalos are much less
common, due to tidal disruption early in the merging process, before
the cluster was virialized. Halos with low relative velocities can
merge shortly after entering the cluster, thus decreasing the number
of small subhalos with low velocities; after virialization, mergers
are suppressed (Gnedin 2003). Large subhalos, $>10^{-3}$ M$_{\rm
cluster}$, are more likely to merge with the central halo, rather than
in subhalo-subhalo mergers (Angulo et al.\ 2008). Thus, in the central
part of the cluster there is also a decrease in the number of massive
subhalos with low velocities.  This may explain why the value of $b$
increases, going from the outer region towards the cluster center. The
baryon content of halos may also affect the velocity bias: Due to
ram-pressure stripping of galactic gas, which is more effective in the
central cluster region (e.g., Arieli, Rephaeli,
\& Norman 2008), the subhalo mass is more easily reduced by tidal
disruption. The deduced subhalo mass function is reduced relative to
that deduced from a corresponding DM-only simulation (Dolag et al.\
2008).

The current data only allow determining an overall typical value of $\beta_{\rm DM}$; 
its best-fit value is in general agreement with the $\beta_{\rm gal}$ deduced from the data, 
and the value of $\beta_{\rm DM}$ found in simulations. As we noted, our assumption of eq.~\ref{sigma_total} 
is likely valid at large radii, which implies that the procedure specified in \textsection~\ref{Methodology} 
is more reliable there.  Host et al.\ (2009) used the X-ray temperature as a surrogate measurement for deriving 
the DM velocity anisotropy profile at intermediate radii by assuming a connection between $T_{\rm gas}$ and 
$T_{\rm DM}$. They applied their analysis to 11 low redshift and 5 intermidiate redshift clusters, including 
A1689. Their deduced values for $\beta_{\rm DM}$ are in very good agreement with our best-fit value. Together 
the two methods enable estimating the DM velocity anisotropy at all radii. Our approach here is less prone to 
the substantial uncertainty inherent in an attempt to determine the DM velocity anisotropy from a similar 
treatment which is based on IC gas properties. A possible difficulty with the latter approach may be due to 
a nonthermal pressure component that could appreciably affect results inferred from a solution to the hydrostatic 
equilibrium equation (e.g., Molnar et al.\ 2010).

It can be seen from figure~\ref{density_profiles_comparison} and
\ref{gamma_comparison} that at $r\sim700$ h$^{-1}$ kpc
$\sim\frac{1}{3}r_{\rm vir}$ there is a change in the relation between
the dynamical properties of the DM and  
galaxies. Carlberg (1994) showed analytically that if 
one assumes 
a power-law mass density profile and isotropic velocity dispersion, 
this yields a power-law profile for the velocity dispersion. He also 
found that in general a cooler tracer has a steeper density profile 
and is more centrally concentrated. We find that at small radii, 
$r\lesssim\ \frac{1}{3} r_{\rm vir}$, $f_{\rm coll}<1$, implying 
that DM is cooler than galaxies. Indeed, at these radii the DM profile is steeper (see 
figure~\ref{gamma_comparison}) and more concentrated (see
figure~\ref{density_profiles_comparison}) than that of 
galaxies. The subhalo distribution would naively be expected to be
closely related to the distribution of galaxies. In fact, galaxies and
subhalos represent different populations and are not directly
comparable since subhalo masses are more strongly affected by tidal
stripping than galactic baryonic matter. It was shown in simulations
that subhalos within $0.3r_{\rm vir}$ typically lose more than 70\% of
their mass during the merging phase, 
while subhalos at $r>0.5r_{\rm vir}$ typically lose only $\lesssim 40\%$ 
of their mass (Nagai \& Kravtsov 2005, hereafter NK05).

It was previously inferred from simulations that there is a spatially
anti-biased subhalo distribution, in the sense that the DM subhalo
profile has a larger core radius than the background DM profile
(DMS04; Gao et al.\ 2004a). DMS04 interpreted their result as due to
missing half of the halo population caused by the known numerical
overmerging problem. However, NK05 used a high-resolution
cosmological cluster simulation to show that the subhalo radial
distribution is significantly less concentrated than that of DM due to
tidal stripping, rather than this being a numerical artifact. They
demonstrated that the radial bias disappears almost entirely if
subhalos are selected using their mass or circular velocity during the
merger phase. A similar result was obtained by Gao et al.\ (2004b)
who found that defining the subhalo population by requiring a minimum
circular velocity gives a subhalo distribution which is more
concentrated than when selection is based on a minimum mass. This
results from the fact that the subhalo distribution is easier to track
by using the maximum circular velocity values, since mass loss is also
accompanied by a decrease in the maximum circular velocity, but the 
decrease is slower than in the mass (NK05).

In many previous studies the observed galaxy distribution was taken to
be cuspy and similar to that of DM (e.g., DMS04; Gao et al.\ 2004a;
NK05), using the galaxy distribution from Carlberg, Yee, \& Ellingson
(1997, hereafter CYE97), who co-added observations of 14 clusters at
various redshifts. The superposed sample contained 1150
galaxies, including background galaxies. 
In the L09 analysis of A1689, 500 cluster members were identified from 
spectroscopic data, and about 1900 cluster members from photometric 
data. L09 showed that the galaxy density profile is best fitted with a 
core profile; although a cuspy profile was found to be acceptable, the 
deduced values of the scale radius and the power-law index were 
questionable. Indeed, the cuspy profile in CYE97 gave a better fit with 
a higher power-law index (preferring Hernquist (1990) rather than an NFW profile), 
yielding a quite high value of the scale radius, $r_s=(0.66\pm
0.09)r_{200}$. Moreover, Adami et al.\ (1998) examined a sample of 62
clusters and found that most galaxy density profiles are better fitted
with a cored rather than a cuspy profile, 
though for individual clusters the preference for a cored profile is 
rarely significant at the 90\% confidence level. When Adami et al.\ 
(1998) composed a superposed sample they obtained 
a clear preference for a core (King) profile (at more than 95 \% 
confidence). Adami et al.\ (1998) showed that CYE97, as well as 
Beers \& Tonry (1986), obtained a cuspy profile due to selection bias 
caused by not taking into account the effect of elongation. In some 
recent simulations (e.g., Saro et al.\ 2006) that included 
gas cooling and star formation, a detailed treatment of stellar 
evolution and chemical enrichment, as well as SN energy feedback in 
the form of galactic winds, the galaxy density profile 
has a shallower core, 
quite different from the cuspy DM profile at small radii, $r<0.4r_{200}$.
 
Finally, we note that our approach of inferring $\beta_{\rm gal}$
using the Jeans equation, which is derived from the collisionless
Boltzmann equation, is not fully self-consistent since at small radii,
$r\lesssim\ \frac{1}{3} r_{\rm vir}$, we find possible deviation from
a fully collisionless behavior. In addition, for simplicity we assumed
that the cluster is spherically symmetric, though there are claims
that this cluster has a significantly triaxial shape (Oguri et al.\
2005; Morandi et al.\ 2010). However, biases due to the triaxial
nature of the cluster should affect the DM and galaxies fairly
similarly since they have a similar distribution. More importantly, it
should be kept in mind that the results reported here are based on a
comprehensive analysis of only one cluster.  Obviously, the results
should be viewed as preliminary until reproduced by a similar analysis
of a sufficiently large sample of clusters. We plan to do so with a
larger sample of relaxed X-ray selected clusters in the CLASH
program\footnote{PI: Marc Postman;
http://www.stsci.edu/$\sim$postman/CLASH/}.

\section*{ACKNOWLEDGMENTS}

We thank Alexey Vikhlinin, Greg Bryan, and Steen Hansen for many
helpful discussions. We acknowledge discussions also 
with Shay Zucker, Ole Host, and Sharon Sadeh. DL acknowledges 
generous support by Dan David Foundation. The work of YR and MN was 
supported in part by US-Israel Binational Science Foundation grant 
452/2008 at Tel Aviv University. RB acknowledges Israel Science Foundation 
grant 823/09. TJB and YR are supported by Israel Science Foundation grant 1218/06


\label{lastpage}


\begin{thebibliography}{9}

\bibitem[Adami et al.(1998)]{1998A&A...336...63A} Adami, C., Mazure, A., Katgert, P., \& Biviano, A.\ 1998, \aap, 336, 63 

\bibitem[Andersson \& Madejski(2004)]{2004ApJ...607..190A} Andersson, K.~E., \& Madejski, G.~M.\ 2004, \apj, 607, 190 

\bibitem[Angulo et al.(2008)]{2008arXiv0810.2177A} Angulo, R.~E., Lacey, C.~G., Baugh, C.~M., \& Frenk, C.~S 2008, arXiv:0810.2177 

\bibitem[Arieli et al.(2008)]{2008ApJ...683L.111A} Arieli, Y., Rephaeli, Y., \& Norman, M.~L.\ 2008, \apjl, 683, L111 

\bibitem[Arnaud et al.(2005)]{2005A&A...441..893A} Arnaud, M., Pointecouteau, E., \& Pratt, G.~W.\ 2005, \aap, 441, 893 

\bibitem[Atrio-Barandela et al.(2008)]{2008ApJ...675L..57A} Atrio-Barandela, F., Kashlinsky, A., Kocevski, D., \& Ebeling, H.\ 2008, \apjl, 675, L57 

\bibitem[Beers \& Tonry(1986)]{1986ApJ...300..557B} Beers, T.~C., \& Tonry, J.~L.\ 1986, \apj, 300, 557 

\bibitem[Benatov et al.(2006)]{2006MNRAS.370..427B} Benatov, L., Rines, K., Natarajan, P., Kravtsov, A., \& Nagai, D.\ 2006, \mnras, 370, 427 

\bibitem[Binney \& Tremaine(1987)]{1987gady.book.....B} Binney, J., \& Tremaine, S.\ 1987, Princeton, NJ, Princeton University Press, 1987, 747 p.,  

\bibitem[Biviano \& Girardi(2003)]{2003ApJ...585..205B} Biviano, A., \& Girardi, M.\ 2003, \apj, 585, 205 

\bibitem[Biviano \& Salucci(2006)]{2006A&A...452...75B} Biviano, A., \& Salucci, P.\ 2006, \aap, 452, 75 

\bibitem[Brada{\v c} et al.(2006)]{2006ApJ...652..937B} Brada{\v c}, M., et al.\ 2006, \apj, 652, 937 

\bibitem[Broadhurst et al.(2005a)]{2005ApJ...621...53B} Broadhurst, T., et al.\ 2005, \apj, 621, 53 (B05a)

\bibitem[Broadhurst et al.(2005b)]{2005ApJ...619L.143B} Broadhurst, T., Takada, M., Umetsu, K., Kong, X., Arimoto, N., Chiba, M., 
\& Futamase, T.\ 2005, \apjl, 619, L143 (B05b)

\bibitem[Broadhurst et al.(2008)]{2008ApJ...685L...9B} Broadhurst, T., Umetsu, K., Medezinski, E., Oguri, M., \& Rephaeli, Y.\ 2008, \apjl, 685, L9 

\bibitem[Bryan \& Norman(1997)]{1997ASPC..123..363B} Bryan, G.~L., \& Norman, M.~L.\ 1997, Computational Astrophysics; 12th Kingston Meeting on Theoretical Astrophysics, 123, 363 

\bibitem[Carlberg(1994)]{1994ApJ...433..468C} Carlberg, R.~G.\ 1994, \apj, 433, 468 

\bibitem[Carlberg et al.(1997)]{1997ApJ...485L..13C} Carlberg, R.~G., et al.\ 1997, \apjl, 485, L13 

\bibitem[Carlberg et al.(1997)]{1997ApJ...478..462C} Carlberg, R.~G., Yee, H.~K.~C., \& Ellingson, E.\ 1997, \apj, 478, 462 (CYE97)

\bibitem[Coe et al.(2010)]{2010arXiv1005.0398C} Coe, D., Benitez, N., Broadhurst, T., Moustakas, L., \& Ford, H.\ 2010, arXiv:1005.0398 

\bibitem[Corless et al.(2009)]{2009MNRAS.393.1235C} Corless, V.~L., King, L.~J., \& Clowe, D.\ 2009, \mnras, 393, 1235 

\bibitem[Clowe et al.(2004)]{2004ApJ...604..596C} Clowe, D., Gonzalez, A., \& Markevitch, M.\ 2004, \apj, 604, 596 

\bibitem[Cole \& Lacey(1996)]{1996MNRAS.281..716C} Cole, S., \& Lacey, C.\ 1996, \mnras, 281, 716 

\bibitem[Col{\'{\i}}n et al.(2000)]{2000ApJ...539..561C} Col{\'{\i}}n, P., Klypin, A.~A., \& Kravtsov, A.~V.\ 2000, \apj, 539, 561   

\bibitem[Diaferio et al.(1999)]{1999MNRAS.307..537D} Diaferio, A., Kauffmann, G., Colberg, J.~M., \& White, S.~D.~M.\ 1999, \mnras, 307, 537 

\bibitem[Diaferio et al.(2005)]{2005ApJ...628L..97D} Diaferio, A., Geller, M.~J., \& Rines, K.~J.\ 2005, \apjl, 628, L97 

\bibitem[Diemand et al.(2004)]{2004MNRAS.352..535D} Diemand, J., Moore, B., \& Stadel, J.\ 2004, \mnras, 352, 535 (DMS04)

\bibitem[Dolag et al.(2008)]{2008arXiv0808.3401D} Dolag, K., Borgani, S., Murante, G., \& Springel, V.\ 2008, arXiv:0808.3401 
 
\bibitem[Ettori et al.(2002)]{2002MNRAS.331..635E} Ettori, S., Fabian, A.~C., Allen, S.~W., \& Johnstone, R.~M.\ 2002, \mnras, 331, 635

\bibitem[Fukushige \& Makino(1997)]{1997ApJ...477L...9F} Fukushige, T., \& Makino, J.\ 1997, \apjl, 477, L9 

\bibitem[Fukushige \& Makino(2001)]{2001ApJ...557..533F} Fukushige, T., \& Makino, J.\ 2001, \apj, 557, 533 

\bibitem[Fukushige \& Makino(2003)]{2003ApJ...588..674F} Fukushige, T., \& Makino, J.\ 2003, \apj, 588, 674

\bibitem[Gao et al.(2004a)]{2004MNRAS.352L...1G} Gao, L., De Lucia, G., White, S.~D.~M., \& Jenkins, A.\ 2004, \mnras, 352, L1 

\bibitem[Gao et al.(2004b)]{2004MNRAS.355..819G} Gao, L., White, S.~D.~M., Jenkins, A., Stoehr, F., \& Springel, V.\ 2004, \mnras, 355, 819 

\bibitem[Ghigna et al.(2000)]{2000ApJ...544..616G} Ghigna, S., Moore, B., Governato, F., Lake, G., Quinn, T., \& Stadel, J.\ 2000, \apj, 544, 616

\bibitem[Gill et al.(2004)]{2004MNRAS.351..410G} Gill, S.~P.~D., Knebe, A., Gibson, B.~K., \& Dopita, M.~A.\ 2004, \mnras, 351, 410

\bibitem[Giodini et al.(2009)]{2009arXiv0904.0448G} Giodini, S., et al.\ 2009, arXiv:0904.0448
 
\bibitem[Gnedin(2003)]{2003ApJ...582..141G} Gnedin, O.~Y.\ 2003, \apj, 582, 141 

\bibitem[Halkola et al.(2006)]{2006MNRAS.372.1425H} Halkola, A., Seitz, S., \& Pannella, M.\ 2006, \mnras, 372, 1425 

\bibitem[Hallman et al.(2007)]{2007ApJ...671...27H} Hallman, E.~J., O'Shea, B.~W., Burns, J.~O., Norman, M.~L., Harkness, R., 
\& Wagner, R.\ 2007, \apj, 671, 27 

\bibitem[Hansen \& Piffaretti(2007)]{2007A&A...476L..37H} Hansen, S.~H., \& Piffaretti, R.\ 2007, \aap, 476, L37 

\bibitem[Hernquist(1990)]{1990ApJ...356..359H} Hernquist, L.\ 1990, \apj, 356, 359 

\bibitem[Host et al.(2009)]{2009ApJ...690..358H} Host, O., Hansen, S.~H., Piffaretti, R., Morandi, A., Ettori, S., Kay, S.~T., 
\& Valdarnini, R.\ 2009, \apj, 690, 358 

\bibitem[Jing \& Suto(2000)]{2000ApJ...529L..69J} Jing, Y.~P., \& Suto, Y.\ 2000, \apjl, 529, L69 

\bibitem[Kawahara et al.(2007)]{2007ApJ...659..257K} Kawahara, H., Suto, Y., Kitayama, T., Sasaki, S., Shimizu, M., Rasia, E., 
\& Dolag, K.\ 2007, \apj, 659, 257 

\bibitem[Kay et al.(2007)]{2007MNRAS.377..317K} Kay, S.~T., da Silva, A.~C., Aghanim, N., Blanchard, A., Liddle, A.~R., Puget, J.-L., Sadat, R., \& Thomas, P.~A.\ 2007, \mnras, 377, 317 

\bibitem[Klypin et al.(2001)]{2001ApJ...554..903K} Klypin, A., Kravtsov, A.~V., Bullock, J.~S., \& Primack, J.~R.\ 2001, \apj, 554, 903

\bibitem[Kravtsov et al.(1998)]{1998ApJ...502...48K} Kravtsov, A.~V., Klypin, A.~A., Bullock, J.~S., \& Primack, J.~R.\ 1998, \apj, 502, 48  

\bibitem[Lemze et al.(2008)]{2008MNRAS.386.1092L} Lemze, D., Barkana, R., Broadhurst, T.~J., \& Rephaeli, Y.\ 2008, \mnras, 386, 1092 

\bibitem[Lemze et al.(2009)]{2009ApJ...701.1336L} Lemze, D., Broadhurst, T., Rephaeli, Y., Barkana, R., \& Umetsu, K.\ 2009, \apj, 701, 1336

\bibitem[Limousin et al.(2007)]{2007ApJ...668..643L} Limousin, M., et al.\ 2007, \apj, 668, 643 

\bibitem[Limousin et al.(2008)]{2008A&A...489...23L} Limousin, M., et al.\ 2008, \aap, 489, 23 

\bibitem[Lu et al.(2006)]{2006MNRAS.368.1931L} Lu, Y., Mo, H.~J., Katz, N., \& Weinberg, M.~D.\ 2006, \mnras, 368, 1931 

\bibitem[Mahdavi(2001)]{2001ApJ...546..812M} Mahdavi, A.\ 2001, \apj, 546, 812
 
\bibitem[Mahdavi et al.(2007)]{2007ApJ...668..806M} Mahdavi, A., Hoekstra, H., Babul, A., Balam, D.~D., \& Capak, P.~L.\ 2007, \apj, 668, 806 

\bibitem[Mamon \& {\L}okas(2005)]{2005MNRAS.363..705M} Mamon, G.~A., \& {\L}okas, E.~L.\ 2005, \mnras, 363, 705

\bibitem[Markevitch et al.(2002)]{2002ApJ...567L..27M} Markevitch, M., Gonzalez, A.~H., David, L., Vikhlinin, A., Murray, S., Forman, W., Jones, C., \& Tucker, W.\ 2002, \apjl, 567, L27
 
\bibitem[Mazzotta et al.(2004)]{2004MNRAS.354...10M} Mazzotta, P., Rasia, E., Moscardini, L., \& Tormen, G.\ 2004, \mnras, 354, 10 

\bibitem[Medezinski et al.(2007)]{2007ApJ...663..717M} Medezinski, E., et al.\ 2007, \apj, 663, 717 

\bibitem[Molnar et al.(2010)]{2010arXiv1002.4691M} Molnar, S.~M., Chiu, I.~-., Umetsu, K., Chen, P., Hearn, N., Broadhurst, T., Bryan, G., 
\& Shang, C.\ 2010, arXiv:1002.4691 

\bibitem[Moore et al.(1999)]{1999MNRAS.310.1147M} Moore, B., Quinn, T., Governato, F., Stadel, J., \& Lake, G.\ 1999, \mnras, 310, 1147 
 
\bibitem[Morandi et al.(2010)]{2010arXiv1001.1656M} Morandi, A., Pedersen, K., \& Limousin, M.\ 2010, arXiv:1001.1656 

\bibitem[Mulchaey \& Zabludoff(1998)]{1998ApJ...496...73M} Mulchaey, J.~S., \& Zabludoff, A.~I.\ 1998, \apj, 496, 73 

\bibitem[Nagai \& Kravtsov(2005)]{2005ApJ...618..557N} Nagai, D., \& Kravtsov, A.~V.\ 2005, \apj, 618, 557 (NK05)

\bibitem[Navarro et al.(1997)]{1997ApJ...490..493N} Navarro, J.~F., Frenk, C.~S., \& White, S.~D.~M.\ 1997, \apj, 490, 493 (NFW)

\bibitem[Navarro et al.(2004)]{2004MNRAS.349.1039N} Navarro, J.~F., et al.\ 2004, \mnras, 349, 1039 

\bibitem[Norman \& Bryan(1999)]{1999ASSL..240...19N} Norman, M.~L., \& Bryan, G.~L.\ 1999, Numerical Astrophysics, 240, 19 

\bibitem[Oguri et al.(2005)]{2005ApJ...632..841O} Oguri, M., Takada, M., Umetsu, K., \& Broadhurst, T.\ 2005, \apj, 632, 841 

\bibitem[Okabe \& Umetsu(2008)]{2008PASJ...60..345O} Okabe, N., \& Umetsu, K.\ 2008, \pasj, 60, 345 

\bibitem[Peng et al.(2009)]{2009ApJ...701.1283P} Peng, E.-H., Andersson, K., Bautz, M.~W., \& Garmire, G.~P.\ 2009, \apj, 701, 1283 

\bibitem[Pointecouteau et al.(2005)]{2005A&A...435....1P} Pointecouteau, E., Arnaud, M., \& Pratt, G.~W.\ 2005, \aap, 435, 1 

\bibitem[Rasia et al.(2004)]{2004MNRAS.351..237R} Rasia, E., Tormen, G., \& Moscardini, L.\ 2004, \mnras, 351, 237 

\bibitem[Ricotti(2003)]{2003MNRAS.344.1237R} Ricotti, M.\ 2003, \mnras, 344, 1237 

\bibitem[Riemer-S{\o}rensen et al.(2009)]{2009ApJ...693.1570R} Riemer-S{\o}rensen, S., Paraficz, D., Ferreira, D.~D.~M., Pedersen, K., 
Limousin, M., \& Dahle, H.\ 2009, \apj, 693, 1570 

\bibitem[Sanderson et al.(2005)]{2005ApJ...630..191S} Sanderson, A.~J.~R., Finoguenov, A., \& Mohr, J.~J.\ 2005, \apj, 630, 191

\bibitem[Saro et al.(2006)]{2006MNRAS.373..397S} Saro, A., Borgani, S., Tornatore, L., Dolag, K., Murante, G., Biviano, A., Calura, F., 
\& Charlot, S.\ 2006, \mnras, 373, 397 

\bibitem[Schmidt \& Allen(2007)]{2007MNRAS.379..209S} Schmidt, R.~W., \& Allen, S.~W.\ 2007, \mnras, 379, 209 

\bibitem[Umetsu \& Broadhurst(2008)]{2008ApJ...684..177U} Umetsu, K., \& Broadhurst, T.\ 2008, \apj, 684, 177 

\bibitem[Umetsu et al.(2009)]{2009ApJ...694.1643U} Umetsu, K., et al.\ 2009, \apj, 694, 1643 

\bibitem[Umetsu et al.(2010)]{2010ApJ...714.1470U} Umetsu, K., Medezinski, E., Broadhurst, T., Zitrin, A., Okabe, N., Hsieh, B.-C., 
\& Molnar, S.~M.\ 2010, \apj, 714, 1470 

\bibitem[Valdarnini(2006)]{2006NewA...12...71V} Valdarnini, R.\ 2006, New Astronomy, 12, 71 

\bibitem[Vikhlinin et al.(2006)]{2006ApJ...640..691V} Vikhlinin, A., Kravtsov, A., Forman, W., Jones, C., Markevitch, M., Murray, S.~S., 
\& Van Speybroeck, L.\ 2006, \apj, 640, 691 

\bibitem[Voit et al.(2005)]{2005MNRAS.364..909V} Voit, G.~M., Kay, S.~T., \& Bryan, G.~L.\ 2005, \mnras, 364, 909 

\bibitem[Wojtak et al.(2005)]{2005MNRAS.361L...1W} Wojtak, R., {\L}okas, E.~L., Gottl{\"o}ber, S., \& Mamon, G.~A.\ 2005, \mnras, 361, L1

\bibitem[Wojtak et al.(2008)]{2008MNRAS.388..815W} Wojtak, R., {\L}okas, E.~L., Mamon, G.~A., Gottl{\"o}ber, S., Klypin, A., 
\& Hoffman, Y.\ 2008, \mnras, 388, 815 

\bibitem[Xue \& Wu(2000)]{2000ApJ...538...65X} Xue, Y.-J., \& Wu, X.-P.\ 2000, \apj, 538, 65 

\bibitem[Xue \& Wu(2002)]{2002ApJ...576..152X} Xue, S.-J., \& Wu, X.-P.\ 2002, \apj, 576, 152 

\bibitem[Zitrin et al.(2009)]{2009MNRAS.396.1985Z} Zitrin, A., et al.\ 2009, \mnras, 396, 1985 

\bibitem[Zitrin et al.(2010)]{2010arXiv1004.4660Z} Zitrin, A., et al.\ 2010, arXiv:1004.4660 


\end{thebibliography}
\end{document}